# Computing Challenges for the Einstein Telescope project

*Stefano* Bagnasco[1*], *Antonella* Bozzi[2], *Tassos* Fragos[3,4], *Alba* Gonzalvez[5], *Steffen* Hahn[6], *Gary* Hemming[2], *Lia* Lavezzi[1], *Paul* Laycock[3,4], *Gonzalo* Merino[7,8], *Silvio* Pardi[9], *Steven* Schramm[4,10], *Achim* Stahl[11], *Andres* Tanasijczuk[12], *Nadia* Tonello[5], *Sara* Vallero[1], *John* Veitch[13], and *Patrice* Verdier[14]

[1]Istituto Nazionale di Fisica Nucleare, Sezione di Torino, I-10125 Torino, Italy
[2]European Gravitational Observatory, I-56021 Cascina (PI) Italy
[3]Départment d'Astronomie, Université de Genève, Chemin Pegasi 51, CH-1290 Versoix, Switzerland
[4]Gravitational Wave Science Center (GWSC), Université de Genève, 24 quai E. Ansermet, CH-1211 Geneva, Switzerland
[5]Barcelona Supercomputing Center, E-08034 Barcelona (Spain)
[6]Institut für Kernphysik (IKP), Karlsruher Institut für Technologie (KIT), D-76344 Eggenstein-Leopoldshafen, Karlsruhe, Germany
[7]Port d'Informació Científica (PIC), Campus UAB, E-08193 Bellaterra (Cerdanyola del Vallès), Spain
[8]Centro de Investigaciones Energéticas, Medioambientales y Tecnológicas (CIEMAT), E-28040 Madrid, Spain
[9]Istituto Nazionale di Fisica Nucleare, Sezione di Napoli, I-80126 Napoli, Italy
[10]Département de Physique Nucléaire et Corpusculaire, Université de Genève, 24 quai E. Ansermet, CH-1211 Geneva, Switzerland
[11]III. Physikalisches Institut, RWTH Aachen University, D-57024 Aachen, Germany
[12]Université Catholique de Louvain, B-1348 Louvain-la-Neuve, Belgium
[13]Institute for Gravitational Research, School of Physics and Astronomy, University of Glasgow, Glasgow G12 8QQ, United Kingdom.
[14]Institut de Physique des 2 Infinis de Lyon - IN2P3, Université Lyon 1, CNRS, F- 69622 Villeurbanne, France

**Abstract.** The discovery of gravitational waves, first observed in September 2015 following the merger of a binary black hole system, has already revolutionised our understanding of the Universe. This was further enhanced in August 2017, when the coalescence of a binary neutron star system was observed both with gravitational waves and a variety of electromagnetic counterparts; this joint observation marked the beginning of gravitational multimessenger astronomy. The Einstein Telescope, a proposed next-generation ground-based gravitational-wave observatory, will dramatically increase the sensitivity to sources: the number of observations of gravitational waves is expected to increase from roughly 100 per year to roughly 100'000 per year, and signals may be visible for hours at a time, given the low frequency cutoff of the planned instrument. This increase in the number of observed events, and the duration with which they are observed, is hugely beneficial to the scientific goals of the community but poses a number of significant computing challenges. Moreover, the currently used computing algorithms do not scale to this new environment, both in terms of the amount of resources required and the speed with which each signal must be characterised. This contribution will discuss the Einstein Telescope's computing challenges, and the activities that are underway to prepare for them. Available computing resources and technologies will greatly evolve in the years ahead, and those working to develop the Einstein Telescope data analysis algorithms will need to take this into account. It will also be important to factor into the initial development of the experiment's computing model the availability of huge parallel HPC systems and ubiquitous Cloud computing; the design of the model will also, for the first time, include the environmental impact as one of the optimisation metrics.

---

[*] Corresponding author: stefano.bagnasco@to.infn.it

# 1 Introduction

Gravitational-wave research is the challenge to detect perturbations that, despite being generated by some of the most energetic events in the Universe, are exceedingly small; this is done by designing interferometers with the highest possible sensitivity and developing algorithms capable of extracting the tiny signal from a large noise background. Gravitational waves are also a key ingredient for multimessenger astronomy in the time domain, one of the most promising fields of research today.

The Einstein Telescope (ET) is the planned European $3^{rd}$ Generation (3G) Gravitational-wave observatory, aiming to build a new infrastructure capable to also host future upgrades for decades without limiting the observation capabilities.

## 1.1 Gravitational-wave signals

In general, gravitational-wave signals can be classified in two broad categories: transient and continuous. Transients can be detected either using coherence methods [1], correlating signals from more than one instrument, or using matched filtering [2] for signals that are well-modelled, and for which signal templates can be reliably generated. The most common transient sources are compact binary coalescences (CBC), while a typical example of an unmodelled burst signal is the explosion of a nearby supernova.

As yet undetected, continuous sources include continuous waves (CW) from, for example, spinning asymmetric systems like stars with an irregularity or asymmetric binaries, and the continuous stochastic background (CGWB), a superimposition of many independent sources that is the gravitational equivalent of the cosmic microwave background.

The detection of signal from each source has different requirements in terms of computing; for example, while the computing power needed for CBC detection and parameter estimation increases significantly with the detector sensitivity, unmodelled burst and continuous sources requirements scale only with the number of instruments and the observation time [3]; or, another example, continuous wave searches are often performed in the frequency domain, so they generally need a sizable amount of data to be collected and pre-processed (Fourier-transformed) before the analysis itself starts.

## 1.2 Computing for gravitational-wave research

In Gravitational-wave research, as indeed in most time-domain astronomy, there exist *three* distinct computing domains.

Data acquisition and preparation, collection of environmental data for detector control, the production of reduced sets of data for different uses and similar tasks comprise the *online* domain.

Following up transient detections with other observatories allows for multi-messenger astronomical analysis of these detections. To enable this, data must be analysed promptly so that alerts can be published in a timely manner. This is the so-called *low-latency* domain: typical tasks are searches, parameter estimation and sky localisation.

Then there is conventional *offline* analysis, in which pre-processed data are usually analysed using conventional batch processing, for example for computing intensive searches or more accurate parameter estimation.

## 2 The Einstein Telescope

ET is designed to reach a sensitivity at least 10 times better than second generation (2G) detectors (the so-called "advanced" detectors aLIGO [4] in the US, AdVirgo [5] in Italy and KAGRA [6] in Japan) on a large fraction of the detection frequency bands, especially in the lower frequency range (from few Hz to 10Hz), where current detectors are essentially blind. Such an increased sensitivity would open a huge potential in astrophysics, fundamental physics and cosmology.

As of today, the ET collaboration comprises 83 research units and more than 1510 members, representing 215 institutions from 24 different countries.

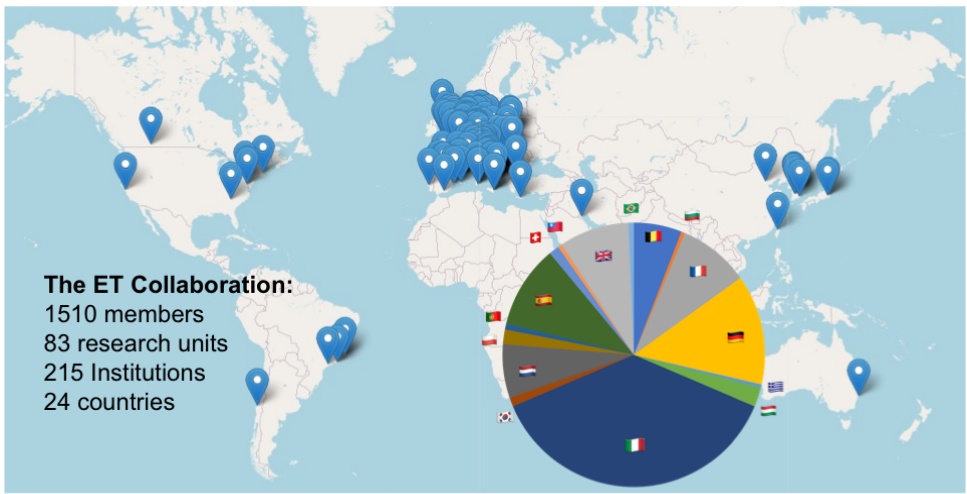

**Fig. 1:** the ET Collaboration

The proposed observatory has been included in the European ESFRI roadmap in June 2021, and the Collaboration was officially formed in June 2022. There are two leading candidates to host the new facility: a site in the EU Regio Rhine-Meuse, close to the Netherlands-Belgium-Germany border, and one close to the Sos Enattos mine in Sardinia, Italy. Site characterization work is ongoing, and a choice is expected to be made by the end of 2024, while the current schedule foresees first data in the 2030s.

### 2.1 Baseline design

In its current baseline design [7], as shown in Fig. 2, the ET has several innovative characteristics:

- it is composed by three detectors in a triangular closed geometry. This allows the use of the "null stream"; by directly summing the signals from the three detectors, any gravitational-wave signal is cancelled. This can be used in several ways in data analysis.
- Each detector consists of two separate Michelson interferometers, one optimised for the high-frequency band and another one, much more challenging, for low-frequency.
- The instrument will be deep underground and cryogenic, to reduce key sources of noise.

Other options involving two separate L-shaped detectors are being currently studied, and many details are not yet finalized; an in-depth discussion of some design choices and their impact is available in [8].

The higher sensitivity means the ability to explore a larger region of the Universe, and up to a point the event rate will scale with the third power of the range at which a given type of event can be detected. So, a tenfold increase in sensitivity translates, very roughly, in three orders of magnitude increase in the detected event rate: a yearly rate of $10^5$ binary neutron star coalescences is expected, and as many binary black holes.

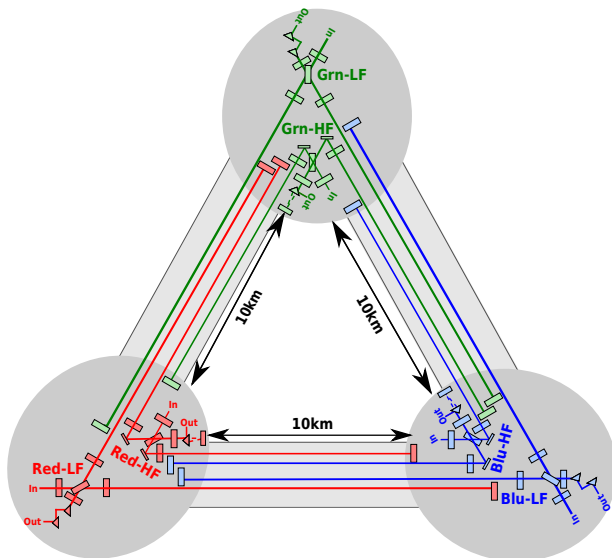

**Fig. 2:** The ET baseline design schematics. Each pair of sides of the triangle contains the arms of two separate interferometers, optimised for higher and lower frequency ranges.

Furthermore, the lower frequency cutoff means that signals stay in the sensitivity band much longer and can be detected earlier. It has already been shown [9] that longer signals like the ones from binary neutron stars can be identified through a dedicated processing in order to issue "early warning" alerts *before* the actual merger occurs; by seconds or minutes in 2G detectors, by hours or even more in 3G ones.

## 2.2 The multimessenger ecosystem

In the 2030's ET is not expected to be observing by itself: another third-generation ground-based interferometer is being proposed in the US, the Cosmic Explorer [10,11]. It is likely that they will be operating as a network, like LIGO, Virgo and KAGRA are today. Also, the

LISA space-based interferometer [12] should be operative, sensitive to very low frequencies and so complementary to ground-based interferometers.

Moreover, several other large electromagnetic and astroparticle facilities will become operational in the same time frame, opening vast possibilities for multimessenger astronomy: optical ground-based telescopes such as the Extremely Large Telescope [13] in Chile; large radio telescope arrays, like the Square Kilometer Array [14] in South Africa and Australia; facilities for cosmic ray astronomy, like the Cherenkov Telescope Array [15] in the Canary Islands and Chile; and neutrino detectors, like the KM3Net [16] in the Mediterranean Sea, just to mention the ones in the ESFRI roadmap. Thus, low-latency data analysis and the prompt exchange of alerts (and possibly of data) will be of paramount importance for multimessenger time-domain astronomy.

**2.3 Einstein Telescope data**

Gravitational-wave data size does not scale with the instruments' sensitivity: the physics channels consist of one time series for each interferometer of the dimensionless "strain" variable, with some associated information on the data quality and detector status. Current 2G detectors produce ~10TB per year each of data for final analysis, so even taking into account the 6-instruments design we expect the yearly ET data to be distributed for final analysis to be a very manageable O(100TB).

Actual raw data, however, do scale with the instrument complexity. These include data from sensors on the instrument monitoring the status of the various systems (vacuum, cryogenics, lasers, seismic attenuators, etc), and environmental data from a very large number of sensors such as seismometers or magnetometers. Current 2G detectors collect auxiliary data from $O(10^5)$ channels, which are used for detector control and characterization or other studies, but not for the final data analysis, and their raw data size is about 1.5PB per year per interferometer (this is the downsampled rate that is written to disk, actual data rate from the detector is much higher), so we expect the raw data size from the ET, that will be stored for safekeeping, to be of the order of some tens of PB per year; while large, this is still not much if compared with other endeavours like collider-based particle physics or radioastronomy.

# 3 Computing challenges

The higher sensitivity of the ET with respect to 2G detectors poses different computing challenges.

First, this implies a huge increase in the expected event rate: as we've seen, a higher sensitivity means the ability to explore a larger portion of the Universe, and an expected yearly rate of $O(10^5)$ CBCs, one every few minutes or more. The need to extract and process this number of candidate events in low-latency, in particular for parameter estimation, means that just scaling current techniques is not enough.

Second, as already mentioned, the lower cutoff frequency implies that events will stay much longer in the sensitivity band of the instrument. This means that longer signal templates (so, more space in memory) and larger template banks will be required to explore a larger part of the parameter space.

Besides devising hierarchical strategies to fully process only the most interesting events, work is needed to improve the algorithms to reduce the computing power: naive scaling of current matched-filter CBC searches and parameter estimation (PE) to 3G design sensitivities will prohibitively require many orders of magnitude more CPU and memory.

Furthermore, the longer signals also have two other consequences that will complicate the analysis, increasing the computing requirements: first, there will often be overlapping signals that need to be disentangled. Second, for signals that stay hours (or even days) in-band, the Earth's movement can no longer be ignored.

The search for continuous waves is currently limited by the available computing power (by orders of magnitude), and not by the detector sensitivity. Like unmodelled burst searches, all-sky CW searches scale linearly with the number of detectors and the observing time, so barring some breakthrough in algorithm development or code optimisation, in 3G detectors they will still be limited by the available computing power [3].

The computing power used to process 2G data today is about 10% of the computing power used for one of the LHC experiments; very preliminary estimates suggest that scaling that power to a few MHS06 years per year, equivalent to 10% of the power used for one LHC experiment in Run 5 (about at the same time of the ET coming online) should be sufficient to process the most important events – i.e., the ones with a good sky localization: we expect O(100) detections per year with a position determined better than 20 deg$^2$ [8] and a high probability of having an electromagnetic counterpart. However, given the mentioned uncertainties in the evolution of the algorithms, having a reliable estimate of the computing needs is a challenge in itself, and will be a task for the coming years.

As for the multimessenger endeavour, many of the producers and consumers of alerts and data will come online before the ET, and several projects such as the International Virtual Observatory Alliance [17] are already designing the upgraded communication infrastructure components: standards, event databases and brokers, automated tools for alert selection, and more are already being developed. The task for the ET is therefore to get involved in the ongoing activities.

## 4 The ET Computing Model

The ET community is already working to define its computing model based on a distributed computing paradigm since the very beginning, taking advantage of the experience of large international experiments and the projects that guarantee the highest level of sustainability and reduced costs.

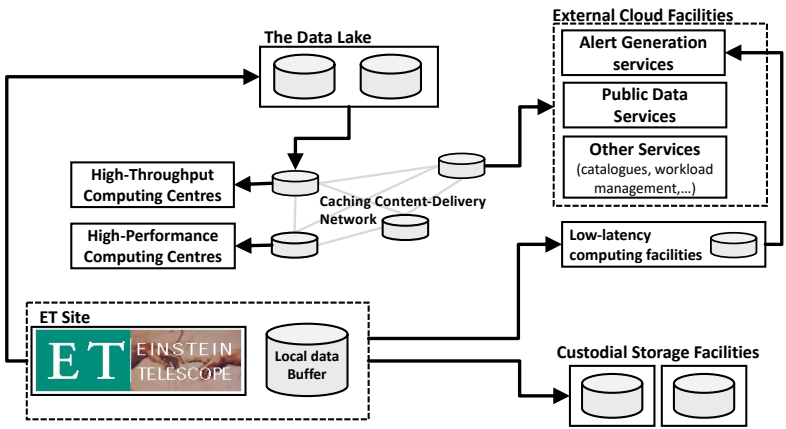

**Fig. 3.** Strawman depiction of ET computing infrastructure (image from the ESFRI proposal)

The starting point is of course the infrastructure currently used by the LVK collaboration, called the International Gravitational-Wave observatories Network (IGWN) [18]. In Fig. 3 a very rough strawman representation of a computing model is shown, based on what is currently used and adapted to some likely evolutions of the European computing infrastructure and some needed changes with respect to 2G computing activities.

As examples, we rely on the existence of a European "Data Lake" to store and distribute the data; we assume many services will run on shared, external Cloud facilities; and we foresee a growing role of High-Performance Computing centres alongside High-Throughput distributed facilities.

It is of course very difficult to foresee exactly what technology, both hardware and software, will be available and what exactly will be the details of the available common infrastructure in the long timescale of the ET. The ET Collaboration has therefore formed, within its e-Infrastructure Board, a Technology Tracking working Group (TTG) to make sure the computing model is kept up to date in terms of the adoption of useful technologies.

For example, while Artificial Intelligence (AI) is not yet a widely used technology in GW data analysis, its role will clearly grow [19]; AI usage may come with the adoption of dedicated architectures for training and inference; the need for on-site High-Performance Computing resources can be driven by the adoption of AI-based techniques for detector control; detector simulation can profit from the development of Digital Twins, and so on.

## 5 ET Mock Data Challenges

The design of the ET computing model will not only provide a blueprint for the infrastructure that will be used to process ET data. Already computing activities are starting to develop the algorithms, simulate the detector performance, and more. We are already deploying a very early prototype of an infrastructure, using existing technologies (currently mostly from the IGWN) to support the upcoming Mock Data Challenge (MDC) activities.

Such MDCs are planned as recurring exercises in which realistic simulated data containing instrumental noise and GW signals, with increasing complexity, are made available to the community with several purposes: engage the gravitational-wave community, develop, optimize, and test data analysis and parameter estimation pipelines, and test the currently available version of the computational infrastructure providing feedback for its evolution.

The plan is to provide as much functionalities as are available at the time of each MDC, and iteratively collect feedback and make the infrastructure evolve to adopt more modern tools and provide even more functionalities tailored to actual community requirements.

## 6 Conclusions

The Einstein Telescope project will realise the European 3$^{rd}$ Generation Gravitational Wave observatory, providing highly enhanced sensitivities to GW detection and enabling a very rich scientific program in astrophysics, fundamental physics, and cosmology.

The high sensitivity of the ET will translate into a high rate of events and a much larger region of the parameter space to explore both for detection and parameter estimation, which in turn will pose many challenges in the fields of computing resource provisioning and exploitation, and algorithms development. The project timescale, however, allows for the pursuit of innovative R&D activities to adapt the computing technologies, algorithms, and strategies to confront these challenges.

The Collaboration is currently in the process of designing its Computing Model and will be building an evolving infrastructure, both in order to cater to the needs of the community from the very beginning and to get feedback on features and performance of the system.